\documentclass{aa}
\usepackage{psfrag,graphicx}
\usepackage{txfonts}
\usepackage{natbib}

\begin{document}

   \title{The Asymmetric Radio Structure and Record Jet \\
          of Giant Quasar 4C\,34.47}
   \subtitle{}
   \author{S. Hocuk \and P.D. Barthel}
   \offprints{P.D. Barthel}
   \institute{Kapteyn Astronomical Institute, University of Groningen,
              P.~O.~Box 800, 9700 AV Groningen, The Netherlands\\
              \email{seyit@astro.rug.nl, pdb@astro.rug.nl}} 
   
   \titlerunning{Giant Quasar 4C\,34.47}   
   \authorrunning{S.~Hocuk \& P.D.~Barthel}
   \date{Received May 21, 2010; accepted August 3, 2010}
   
\abstract
   {AGN unification models predict that all radio sources associated with QSOs should
    make a substantial angle with the sky plane.}
   {Using the morphological as well as polarization properties of a giant radio source
    associated with a QSO, the predictions of the orientation unification model are
    tested.}
   {Giant double-lobed radio source 4C\,34.47, associated with QSO B1721+343 is
    mapped at arcsecond scale resolution, and the data are subsequently analyzed
    within the context of current models for extragalactic radio sources.}
   {4C\,34.47 displays a straight one-sided jet, measuring a record length of
   380\,kpc, in its double-lobed radio structure. Assuming an intrinsically
   symmetric two-sided jet structure the radio source jet axis must be at least
   33$^{\circ}$ away from the sky plane, that is within 57$^{\circ}$ from the
   line of sight. The radio polarization properties
   indicate that this giant source has largely outgrown the depolarizing halo
   generally associated with the host galaxies of powerful radio sources.
   The measured small depolarization asymmetry is nevertheless in accordance
   with its inferred orientation.}
   {All data for this giant radio source are in agreement with its
    preferred orientation as predicted within the unification scheme
    for powerful radio sources. Seen under a small aspect angle the 
    radio source is large but not excessively large. The global properties
    of 4C\,34.47 do not differ from other giant (old) FR2 radio sources: 
    it is a slowly expanding low-luminosity radio source.}
   
\keywords{Galaxies: active -- quasars: individual: 4C\,34.47} 

\maketitle

\section{Introduction}

Quasars were discovered in the 1960s by virtue of their radio emission. 
While this emission later proved to be exception rather than the rule,
the radio sources associated with quasi-stellar objects have found great
interest over the past decades.  With ever increasing angular resolution
(including VLBI), the study of extragalactic radio sources -- quasars
and radio galaxies -- led to our understanding of the relevant
processes, whereby the ultimate energy source in these objects must be
located in their nuclei, on the subparsec scale, and whereby jets
transport some part of this energy to radio-emitting lobes in intra- or
intergalactic space.  In fact, the study of extragalactic radio sources
and the subsequent need for increasing angular resolution provided the
stimulus to develop VLBI techniques. 

Fine reviews of the morphologies and other properties of extragalactic
radio sources can be found in e.g., \citet{Miley1980},
\citet{Bridle1984}, \citet{Saikia1988}, \citet{Carilli1996}, and
\citet{Zensus1997}.  The radio morphologies of quasars can be broadly
classified into two categories: core-dominated or lobe-dominated. 
Examination of the detailed properties of these two classes,
incorporating properties in other wavelength regimes, has led to the
notion that the orientation of the radio jets with respect to the
observer is the prime discriminant.  Core-dominated objects typically
display one short (curved) jet and relatively weak halo-emission besides
the dominant core (coincident with the actual QSO), whereas the
morphologies of lobe-dominated (i.e., Fanaroff \& Riley Class 2, FR2)
objects comprise extended double lobes straddling the QSO radio core. 
Only single-sided jets are observed, in either classes.  This jet
asymmetry together with the occurrence and the magnitude of the
milliarcsec scale proper motion measured in the jet components led to
the picture whereby radio-loud quasars are preferably oriented objects,
the lobe-dominated objects being at substantial angles to the
line-of-sight, the core-dominated objects being at small angles to that
line.  Projection, foreshortening, and relativistic effects are
naturally invoked to explain the observed radio properties of the
compact subclass, including the frequently observed superluminal motion
\citep[e.g.,][]{Orr1982}.  The combined properties of the radio-loud QSO
class together, however, imply that quasars in or close to the plane of
the sky simply do not exist: such objects must masquerade as another
class of object, namely extended double-lobed FR2 radio galaxies. 
Following the spectropolarimetric detection of an obscured quasar in
radio galaxy 3C\,234 \citep{Ski1984}, this unified picture for powerful
extragalactic radio sources was initially explored by
\citet{Scheuer1987} and \citet{Peacock1987}, and subsequently put on a
firm basis by \citet{Barthel1989b}.  An optically opaque `torus'
\footnote{at least a dust configuration covering a substantial solid
angle} surrounding the central accretion disk in its equatorial plane,
blocking the view towards the accretion disk and its associated broad
line region, is an essential ingredient of these unification models. 
Relevant reviews of the quasar -- radio galaxy as well as complementary
unification theories can be found in \citet{Antonucci1993},
\citet{Urry1995}, and \citet{Maiolino2002}.  We stress that a complete
explanation of extragalactic radio sources should obviously also take
into account the effects of radio source evolution
(youth-adulthood-seniority), radio source environment, and activity duty
cycle, besides the aspect-dependent effects. 

Their asymmetric radio depolarization (``depolarization asymmetry'')
provided independent support for the preferred orientation of the quasar
class.  Following up on radio galaxy depolarization studies by e.g.,
\citet{Strom1988}, \citet{Laing1988} and \citet{Garrington1988} drew
first attention to the fact that the ``jetted'' lobes in quasars display
less depolarization (with increasing wavelength) than the unjetted
lobes.  They suggested that this effect was due to natural line-of-sight
depolarization effects towards a near lobe with approaching jet and a
distant lobe with receding jet, through an extended, Faraday-thick,
magneto-ionic halo hosting the radio source.  The effect appeared to be
strong in small radio sources, but virtually absent in large ones
\citep{Garrington1991b}. 

Extensive radio imaging studies by e.g., \citet{Hough1989},
\citet{Hough1999}, \citet{Bridle1994}, \citet{Fernini1993, Fernini1997},
\citet{Dennett1997, Dennett1999} and most recently \citet{Mullin2008}
have yielded broad agreement with the orientation-dependent unification
scheme, but the abovementioned studies also found evidence for radio
source environmental effects.  Consistency with aspect-dependent
unification was also found in optical and infrared studies of the AGN
and their host galaxies \citep[e.g.,][]{Ogle1997, Dunlop2003, Haas2004,
Haas2005}.  Open issues obviously remain, such as the torus opening
angle \citep[e.g.,][]{Grimes2004}, the nature of the class of Broad Line
Radio Galaxies \citep[e.g.,][]{Dennett2000, Bemmel2001}, and the nature
of Low Excitation Radio Galaxies, LERGs \citep[e.g.,][]{Hardcastle2004}
as well as mid-infrared weak radio galaxies \citep[e.g.,][]{Ogle2006,
Wolk2010}

For more than a decade, the 4-arcmin radio source 4C\,34.47, associated
with the $z=0.206$ QSO 1721+343 (B1950), was known as the largest quasar
\citep{Conway1977, Jagers1982}.  Using a flat cosmology, with $H_0 =73$
and $\Omega_m = 0.27$\footnote{values used throughout}, its projected
linear size is 0.84~Mpc (the scaling factor is 3.25 kpc/arcsec).  That
size record was taken over in 1989 by 4C\,74.26 \citep{Riley1989},
measuring 1.1~Mpc.  Having a projected dimension of 2.3~Mpc, WENSS
B\,0750+434 is the current record holder, although its discoverers
\citep{Schoen2001} did not seem to be aware of the fact that their giant
quasar is substantially larger than the then current record holder
HE\,1127$-$1304 \citep{Bhat1998}. 

With the aim to study the detailed properties of 4C\,34.47, a series of
multi-frequency VLA observations was conducted in the mid 1980-s (co-I's
Barthel, van Breugel, J\"agers).  Some initial radio images were
obtained and published \citep{Barthel1987} but the full analysis of the
radio morphological, spectral, and polarization properties was never
completed.  In the meantime, the bright radio core of 4C\,34.47 was
targeted by successive series of VLBI observations, from 1980 onwards
\citep{Breugel1981, Barthel1985, Barthel1989a, Hooi1992}.  The detection
of superluminal motion in the core of this giant radio source
\citep{Barthel1989a} obviously provided substantial support for the
preferred orientation of the radio-loud quasar class. 

The book not being closed on this remarkable object combined with
renewed interest in giant radio sources, within both radio source
unification and evolution models \citep[e.g.,][]{Ishwara1999,
Machalski2006} led us to reexamine the original VLA data.  Do its
large-scale radio properties agree with the proposition that 4C\,34.47
-- despite its giant projected dimension -- is oriented relatively close
to the line of sight, as inferred from its nuclear radio properties?
This is the prime question to be addressed in this paper. 

\section{Observations, Data Reduction and Imaging}

\subsection{Observations}

Radio observations of 4C\,34.47 have been made with the VLA using its
C-band (6cm) and L-band (20cm) and at three different array
configurations, yielding four data sets: two different resolutions at
two different wavelengths.  All data were taken in 1984 by Barthel and
Van Breugel.  Typical resolution parameters are listed in Table
\ref{tab:table1}. 

\begin{table}[htb]
    \caption{Typical VLA resolutions $\theta_{HPBW}$ in arcseconds, in
     the current project. The longest baselines corresponding to each 
     of the configurations are 10\,km for B, 3.6\,km for C and 1\,km for D.}
\begin{center}
\begin{tabular}{|c|c|c|}
\hline
VLA Configuration & C band & L band \\
\hline
\hline
B                 &   n/a  &  3{\farcs}9  \\
C                 &   3{\farcs}9 & 14\arcsec \\
D                 &   14\arcsec  &  n/a   \\
\hline
\end{tabular}
    \label{tab:table1}
\end{center}
\end{table}

On-source integration times for the observations are in the range of 1.5
to 3 hours for each configuration/band combination.  10 to 20 minutes
scans on 4C\,34.47 covering a range of hour angles were interspersed
with short observations of nearby phase and amplitude calibrator
B1732+389.  3C\,286 served as absolute amplitude calibrator and also to
correct for the phase difference of right and left polarization. 
Coordinates of the radio sources are shown in Table \ref{tab:table2}. 
Two intermediate frequencies (IFs) with bandwidths ranging from
12.5\,MHz to 50\,MHz were used.  The detailed observing parameters are
listed in Table \ref{tab:table3}. 

\begin{table}[htb]
\caption{Positions of the target and calibration sources in 1950 (FK4) coordinates.}
\begin{center}
\begin{tabular}{|l|l|l|}
\hline
Source name      & RA                                & Dec                        \\
\hline
\hline
4C\,34.47 & 17$^{\rm h}$ 21$^{\rm m}$ 32{\fs}02   & 34\degr 20\arcmin 41{\farcs}4   \\
3C\,286   & 13$^{\rm h}$ 28$^{\rm m}$ 49{\fs}6577 & 30\degr 45\arcmin 58{\farcs}640 \\ 
1732+389  & 17$^{\rm h}$ 32$^{\rm m}$ 40{\fs}4875 & 38\degr 59\arcmin 46{\farcs}932 \\
\hline
\end{tabular}
    \label{tab:table2}
\end{center}
\end{table}

\begin{table}[htb]
    \caption{Observation details, for the different configurations. Code name LB
      indicates L-band B-array, and so forth}
\begin{center}
\begin{tabular}{|c|c|l|}
\hline
Observation code name         & Time       & Epoch \\
                              & on source  &       \\
                              & (minutes)  &       \\
\hline
\hline
LB     & 107 & 1984 Jan.20   \\
LC     & 119 & 1984 Apr.09   \\
CC     & 191 & 1984 Apr.11   \\
CD     & 93  & 1984 Jul.29   \\
\hline
\end{tabular}
    \label{tab:table3}
\end{center}
\end{table}

\subsection{Data Reduction}

The array performed well: judged from the calibration sources, the
antenna phase, amplitude, and polarization calibration appeared stable
to within a few percent.  The radio data were of high quality, and there
was no need for extensive flagging of discrepant points.  Reduction of
the data was performed using standard NRAO AIPS image processing
routines, including several steps of self-calibration (phase only,
followed by amplitude self-calibration).  Several successive
self-calibration and cleaning cycles with varying amplitude gain factors
generally led to a rapid convergence towards the final images.
In order to uncover the weak diffuse emission, the final deep cleaning
steps involve well over 100.000 iterations.

Multi-resolution images were obtained by combining $uv$-datasets. 
However, 4C\,34.47 is known to have a (strong) variable radio core. 
Hence the multi-resolution images involve subtraction of the cores in
both data sets, with subsequent calibrating, imaging and cleaning of the
concatenated data set, and restoration of a core in the resulting
multi-resolution image. 

\section{Results}

\subsection{High and low resolution images: a record jet in a
            low luminosity radio source}

\begin{figure*}[htb!]
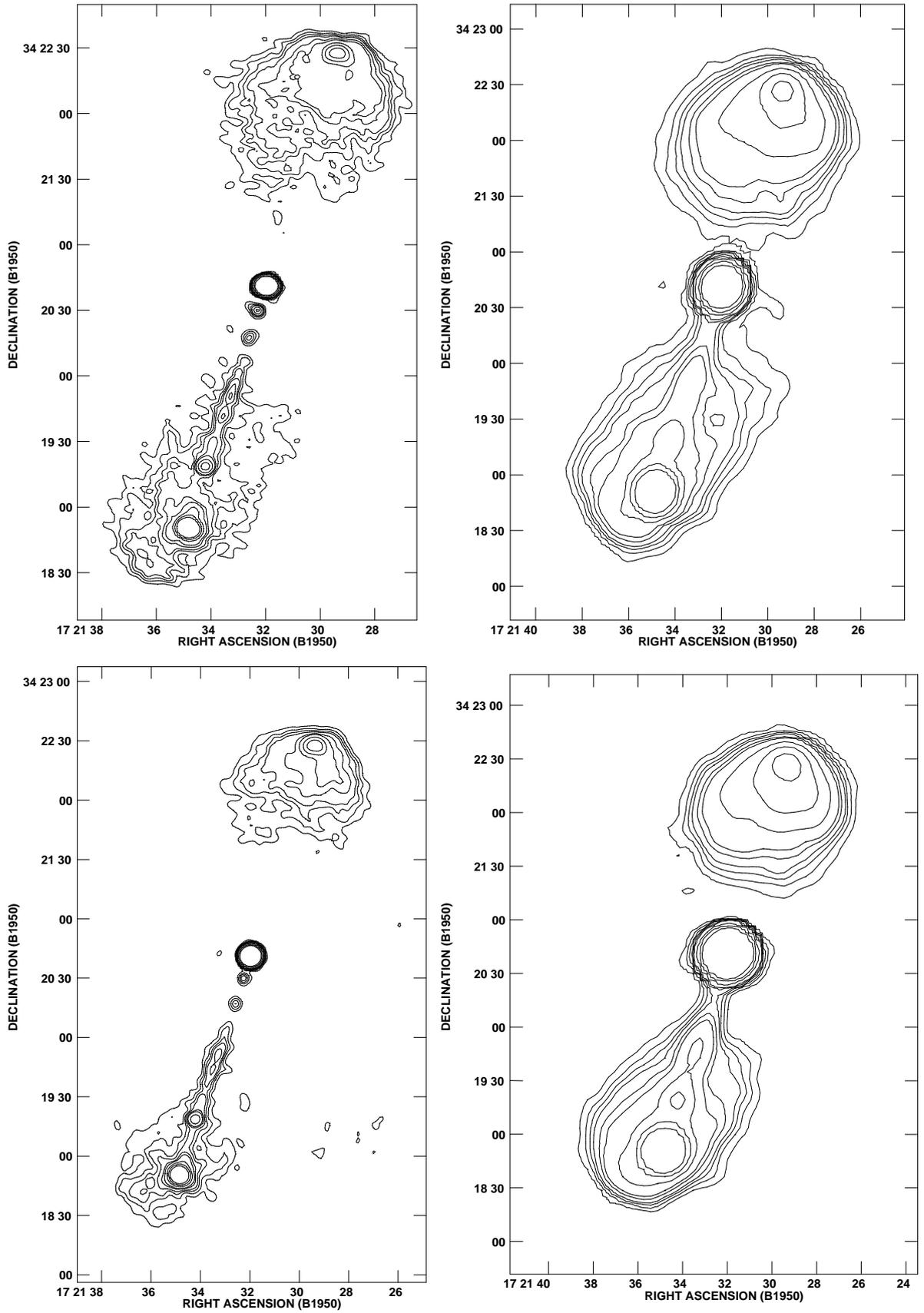

\centering
\begin{tabular}{l l}
\begin{minipage}{7cm}
\includegraphics[scale=0.45]{New-A1-3+sigma2.eps}
\end{minipage} &

\begin{minipage}{7cm}
\includegraphics[scale=0.45]{New-A3-3+sigma2.eps}
\end{minipage} \\

\begin{minipage}{7cm}
\includegraphics[scale=0.45]{New-A4-3+sigma2.eps}
\end{minipage} &

\begin{minipage}{7cm}
\includegraphics[scale=0.45]{New-C12-3+sigma2.eps}
\end{minipage} \\

\end{tabular}
\caption{Contour plots of the four images. The contour levels are RMS 
 (table (\ref{tab:table4}) $\times$ 3, 6, 9, 12, 18, 24, 48, 96, 192 mJy Beam$^{-1}$. 
 Top left: high resolution 20cm image (LB). Top right: low resolution 20cm image (LC).
 Bottom left: high resolution 6cm image (CC). 
 Bottom right: low resolution 6cm image (CD).}
\label{fig:figure1}
\end{figure*}

The final total intensity images are shown in Fig. \ref{fig:figure1}. 
A codename (band+array) is given to each image cf. Table
\ref{tab:table3}.  The high resolution images (LB and CC) show details
of the jets, while in the relatively low resolution images (LC and CD)
the diffuse lobe structures are better seen.  4C\,34.47 has a remarkably
bright core, unlike typical double-lobed 3CR quasars whose cores
generally contribute $\la 10\%$ of the total 6 or 20 cm flux density
\citep[e.g.,][]{Bridle1994}.  Table \ref{tab:table4} summarizes the
resulting imaging figures.  While the theoretical noise was not reached,
acceptable figures were obtained.  The resulting angular resolution is
$\sim$4\arcsec~ and $\sim$12\arcsec~ respectively, cf.  expectation. 
Given that the overall source angular size is just over 4\arcmin,
primary beam corrections are not necessary (the VLA 6cm primary beam
width $R \simeq 1.22 \cdot \lambda / D \simeq 9{\farcm}8$). 

\begin{table}[htb]
\caption{Observational parameters.}
\begin{center}
\begin{tabular}{|l|l|l|l|l|}
\hline
Code & Freq.  & Beam size      & Peak flux d. & RMS noise\\
name & (GHz)  & $\theta_{maj}$, $\theta_{min}$, PA & (mJy/beam) & (mJy/beam)      \\
\hline
\hline
LB  & 1.47115 & 4.60, 3.83,    83{\fdg}16  & 477.8      & 0.08 \\ 
LC  & 1.47740 & 12.56, 11.72, --5{\fdg}75  & 489.3      & 0.11 \\ 
CC  & 4.84760 & 4.75, 4.61,    83{\fdg}57  & 339.5      & 0.04 \\ 
CD  & 4.86010 & 12.96, 12.83,  51{\fdg}26  & 337.2      & 0.05 \\ 
\hline
\end{tabular}
    \label{tab:table4}
\end{center}
\end{table}

The images show that both hot spots and the core are well aligned, to
better than 1$^{\circ}$.  The high resolution 6cm image (CC) displays a
beautiful straight one-sided jet, which stretches almost continously
from the core to the southern hot spot -- a (projected) distance of
380\,kpc.  For comparison, the straight part of the NGC\,315 jet
measures 310\,kpc \citep[e.g.,][]{Bridle1979, Mack1997}.  To our best
knowledge, the jet in 4C\,34.47 is the record longest straight jet. 
Even at relativistic speeds, the particle travel time from core to
southern hot spot well exceeds one million years.  This southern hot
spot is more compact than the northern one; it is also more distant from
the core than the northern one is.  No counterjet (towards the northern
lobe) is seen.  As alluded to already, the intensity of the core is
high: its relative contribution is $\sim$35\% of the 20cm and $\sim$58\%
of the 6cm total intensity.  Its spectral index $\alpha^{5GHz}_{1.4GHz}
= -0.3$ (S$_{\nu} \propto \nu^{\alpha}$).  From the low resolution data
we infer a total 1.4\,GHz flux density of 1.4\,Jy; taking out the
probably relativistically boosted core emission then yields a 1.4\,GHz
integrated radio source luminosity of $1.1 \times 10^{26}$ W/Hz.  With
reference to \citet{Fernini1991}, we note the resemblance of 4C\,34.47
to FR2 quasar 3C\,47, concerning their overall radio morphologies and
the knot structure in their one-sided jets (but we keep in mind that the
latter is roughly a factor twenty more radio-luminous than the former). 
 
\subsection{Multi-resolution combined images}

\begin{figure*}[htb!]
\begin{tabular}{l l}

    \begin{minipage}{8.8cm}
    \includegraphics[scale=0.44]{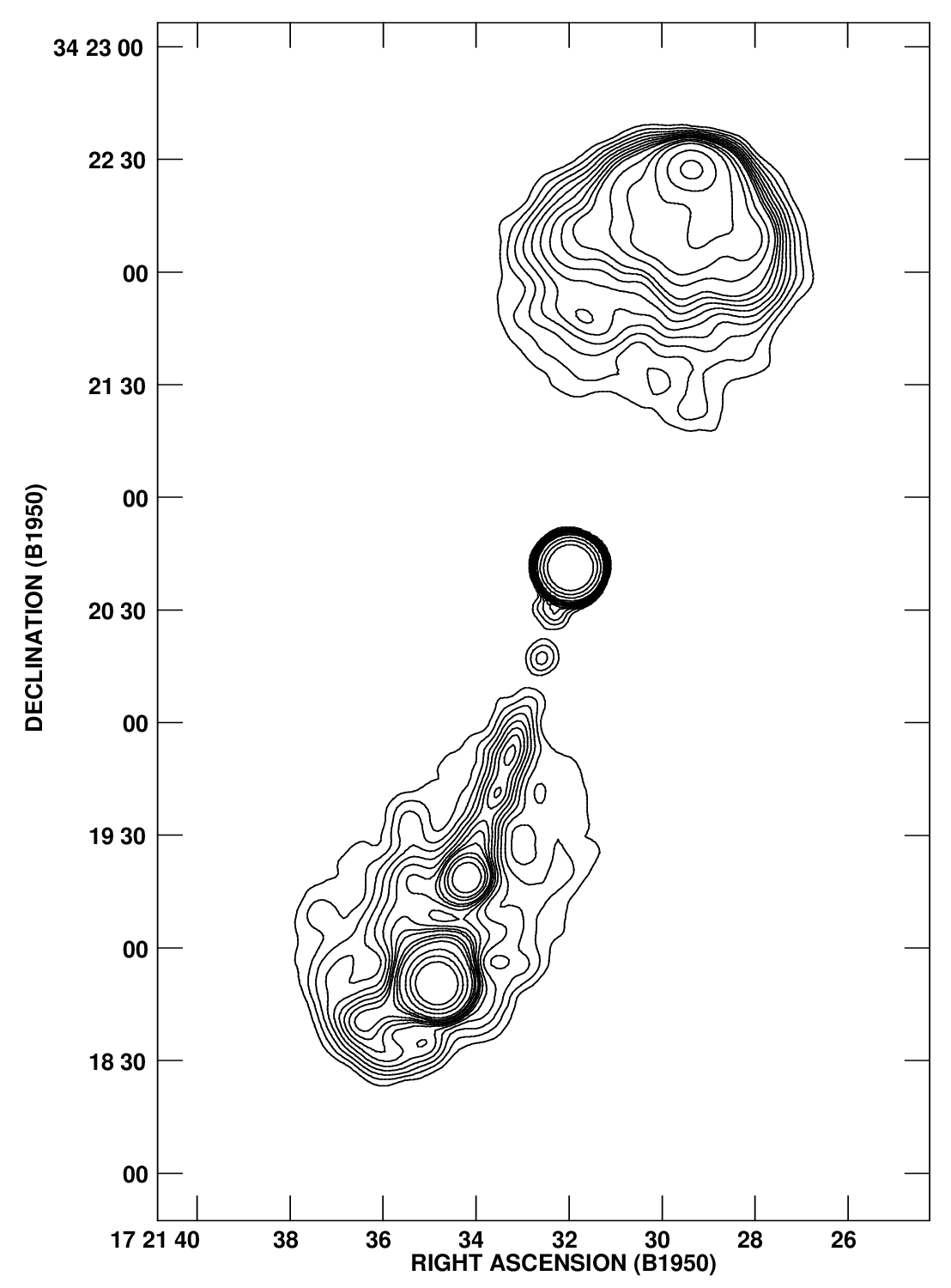}
    \caption{Combined 1.4\,GHz image of two different resolution maps,
     LB and LC (see Fig. 1). The FWHM of the clean beam was fixed at 7{\farcs}5.
     Contour levels are 0.25 $\times$ 4, 6, 8, 10, 12, 14, 16, 18, 20, 25, 30, 40,
     80, 160, 320 mJy/beam. The peak flux density is 487.7 mJy/beam.}
    \label{fig:figure2}
    \end{minipage} &

    \begin{minipage}{8.8cm}
    \includegraphics[scale=0.44]{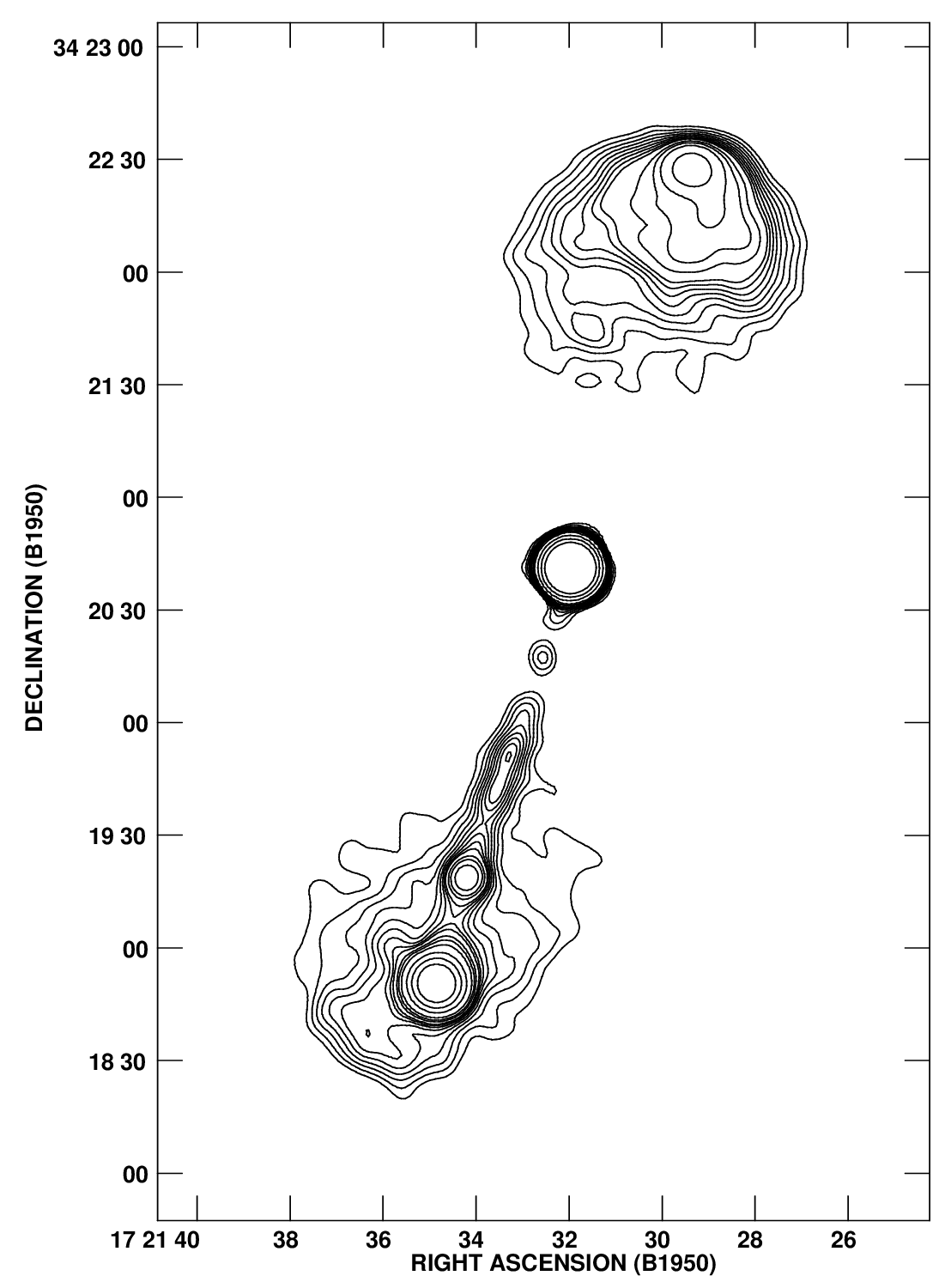}
    \caption{Combined 5\,GHz image of two different resolution maps,
     CC and CD (see Fig. 1). The FWHM of the clean beam was fixed at 7{\farcs}5.
     Contour levels are 0.109 $\times$ 4, 6, 8,
     10, 12, 14, 16, 18, 20, 25, 30, 40, 80, 160, 320 mJy/beam.
     The peak flux density is 337.2 mJy/beam.}
    \label{fig:figure3}
    \end{minipage}

\end{tabular}
\end{figure*}

Images at the same frequency but with different resolutions can be
combined to obtain a multi-resolution image.  These then show both the
detailed features of the high resolution and the low surface brightness
features of the low resolution image.  Employing core subtraction as
described above, we obtained such images at 20cm and at 6cm; these are
shown in Fig.\ref{fig:figure2} (LB+LC) and Fig.\ref{fig:figure3}
(CC+CD), respectively.  A fixed circular clean beam of 7{\farcs}5 was
adopted, in order to facilitate intercomparison of the images, incl. 
radio spectral behaviour.  The images reach noise levels of $\sim 0.12$
and $\sim 0.06$ mJy/beam respectively. 

The straight jet stands out nicely in the 6cm image.  It displays a
number of knots and connects to the more distant of the two hot spots, cq.
radio lobes. No counter-jet is observed. 

\subsection{Core variability}

The variability of the radio core is noteworthy.  We compare the
measured 1984 core strength with earlier measurements made with the
Westerbork Synthesis Radio Telescope (WSRT), in 1973/74
\citep{Conway1977} and in 1974/79 \citep{Jagers1982}.  The results are
summarized in Table \ref{tab:table5}.  The data indicate a $\sim$15\%
decrease in ten years at 1.4\,GHz, and $\sim$25\% at 5\,GHz.  It should be
noted that the associated optical QSO 1721+343 was also found to be
variable \citep{McGimsey1978}.  Relativistic beaming effects, invoked to
explain the measured superluminal motion \citep{Barthel1989a, Hooi1992},
provide a natural explanation for the behaviour of the (non-thermal)
core emission. 

\begin{table}[htb!]
\caption{4C\,34.47 core strength comparison}
\begin{center}
\begin{tabular}{|c|c|c|c|}
\hline
Epoch  & 20cm flux d. & 6cm flux d. & reference \\
\hline
\hline
1973.9  & 580$\pm$20 mJy  & ---             & Conway et al. \\
1974.3  & ---             & 508$\pm$20 mJy  & Conway et al. \\
1976.9  & 610$\pm$30 mJy  & ---             & J\"agers et al. \\
1979.3  & ---             & 440$\pm$30 mJy  & J\"agers et al. \\
1984.1  & 480$\pm$6.6 mJy & ---             & this paper  \\
1984.3  & 499$\pm$8.6 mJy & 340$\pm$3.4 mJy & this paper  \\
1984.6  & ---             & 339$\pm$5.5 mJy & this paper  \\
\hline
\end{tabular}
    \label{tab:table5}
\end{center}
\end{table}

\subsection{Spectral index behaviour}

\begin{figure*}[htb!]

\begin{minipage}[c]{19cm}
\begin{center}
\begin{tabular}{|c|c|c|c|c|c|c|c|c|c|}
\hline
Position & A & B & C & D & E & F & G & H & I \\
\hline
\hline
Spectral index $\alpha$ & $-0.86$ & $-0.77$ & $-0.29$ & $-0.75$ & $-0.50$ & $-0.42$ & $-0.79$ & $-0.84$ & $-0.88$ \\
\hline
\end{tabular}
    \label{tab:table6}
\end{center}
\end{minipage} 

\begin{minipage}[c]{18cm}
    \centering
    \includegraphics[scale=0.5]{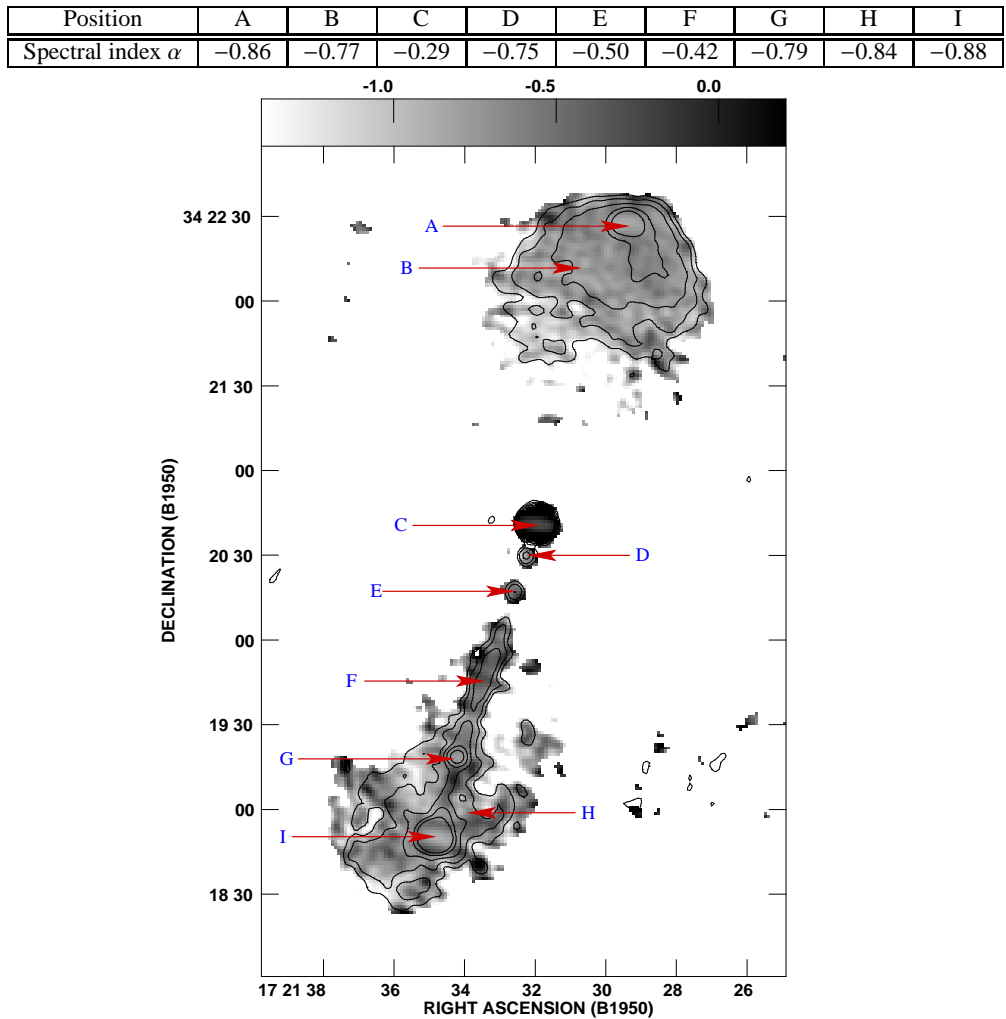}
    \caption{Gray scale representation of the spectral index between 1.4 and 5\,GHz
     with superimposed contours of total intensity. The measurements at the marked
     locations are summarized in the table.}
    \label{fig:figure4}
\end{minipage}

\end{figure*}

Fig.  \ref{fig:figure4} presents the spectral index map
($\alpha^{5GHz}_{1.4GHz}$) for 4C\,34.47, taken from its high resolution
images, with capital letters marking positions of interest.  Both hot
spots (positions A and I) display steep emission, with values $-0.86$
and $-0.88$, respectively.  Lobe emission is also steep, although not
extreme: we measure indices $-0.77$ and $-0.84$ at locations B and H. 
As alluded to before, the core (C) has a fairly flat index $-0.29$; jet
knots display values between $-0.42$ to $-0.79$ (D to G).  The spectral
index error is estimated to be $\pm0.05$ at most.  A summary of the
measurements is given in the Table above the spectral index map. 

When comparing the measured spectral indices with the values obtained by
\citet{Jagers1982}, which are $\sim -0.75$ at the hot spots and $\sim
-0.25$ in the center, we see that they are in good agreement.  Only
slight differences exist which can be explained by their larger spectral
range (0.6\,GHz to 5\,GHz) and by core variability (their core spectral
index is flatter). 

\subsection{Polarization properties}

The polarization properties of 4C\,34.47 were examined by analyzing the
polarized intensity, degree of linear polarization and polarization
angle maps of the radio source.  These maps were obtained by combining
I, Q and U images, which describe the polarization parameters $P$
(polarized intensity), $\Pi_{L}$ (polarization percentage), and
$\varphi$ (polarization angle (E-vector)), according to Eqs.
(\ref{eq:eq1}) through (\ref{eq:eq3}):

\begin{equation}
P = C \cdot \sqrt{Q^2 + U^2}
\label{eq:eq1}
\end{equation}

\begin{equation}
\Pi_{L} = \frac {P} {I}
\label{eq:eq2}
\end{equation}

\begin{equation}
\varphi = \frac {1} {2} tan^{-1} \frac {U} {Q}
\label{eq:eq3},
\end{equation}

\noindent
where the factor C is a noise-based correction for Ricean bias.  Figs.
5 and 6 present the polarization data, for both IFs at both frequencies
in the low resolution (14\arcsec) configuration, superposed on total
intensity images. 

Substantial Faraday rotation is seen between the L-band and the C-band
data, the nature of which will be discussed below.  Both the polarized
intensity (Fig.~5) and the polarization percentage, P/I (Fig.~6) are
shown.  Little depolarization is seen.  We measure typical polarization
percentage values of $\sim$ 10\% -- 20\% in the jet and inner lobe
regions, at both frequencies.  The lobe edges, particularly in the
southern lobe, display polarization percentages $\sim$ 25\% -- 40\%. 
Assuming that the polarization vectors at C-band trace the source
intrinsic fields (see below), the well ordered magnetic fields follow
the edges of the lobes.  Intricate field structure is seen in the inner
lobe regions.  The jet magnetic field is parallel to the flow but
experiences a sudden turn in the bright jet knot, 30\arcsec~ (100\,kpc)
before the southern hot spot.  This is also seen in the high resolution
C-band polarization image (not shown here). \\

\begin{figure*}[htb!]
\centering
\begin{tabular}{l l}
\begin{minipage}{7cm}
\psfrag{RIGHT ASCENSION \(B1950\)}{}
\includegraphics[scale=0.45]{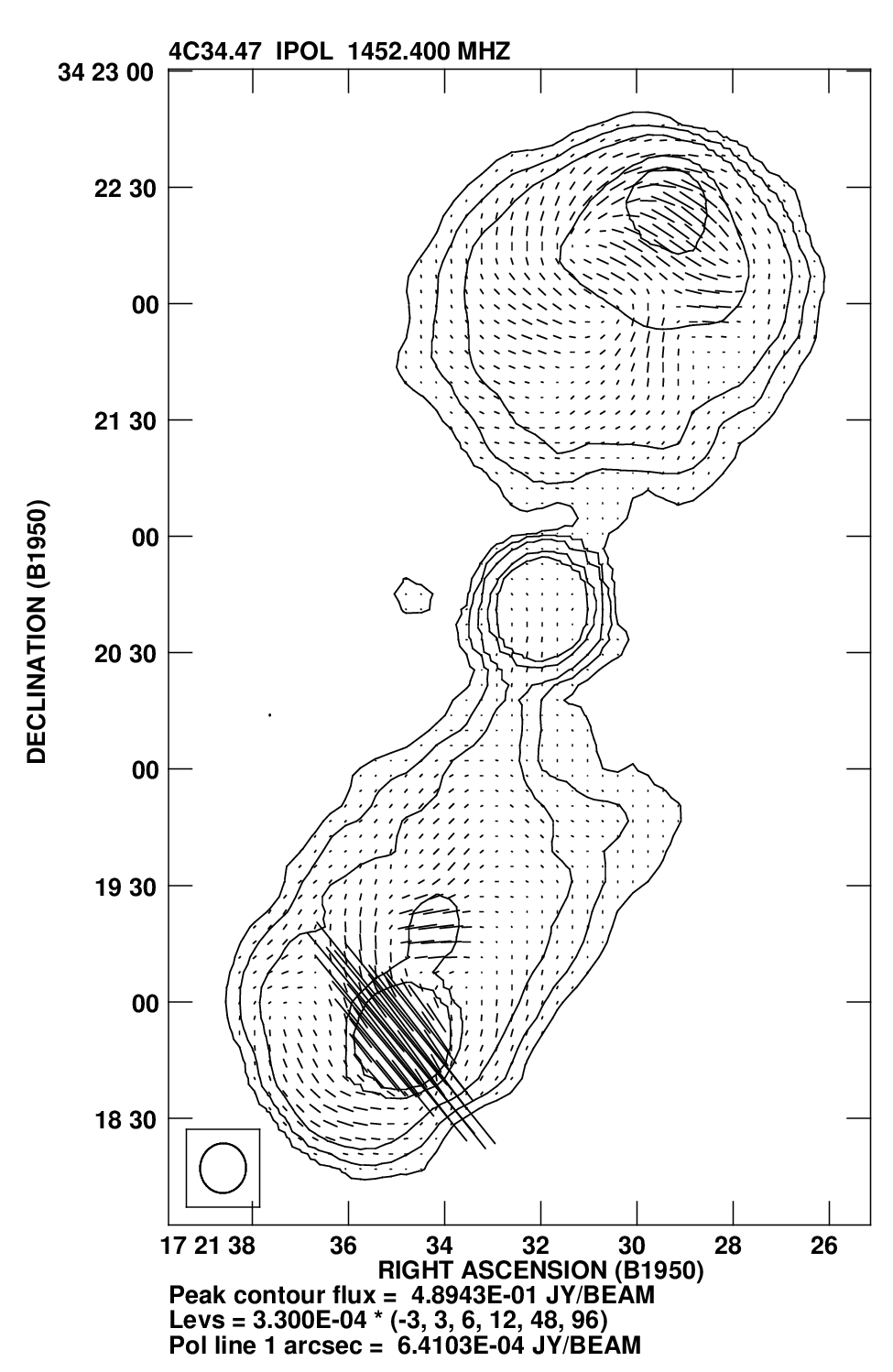}
\end{minipage} &

\begin{minipage}{7cm}
\psfrag{RIGHT ASCENSION \(B1950\)}{}
\psfrag{DECLINATION \(B1950\)}{}
\includegraphics[scale=0.45]{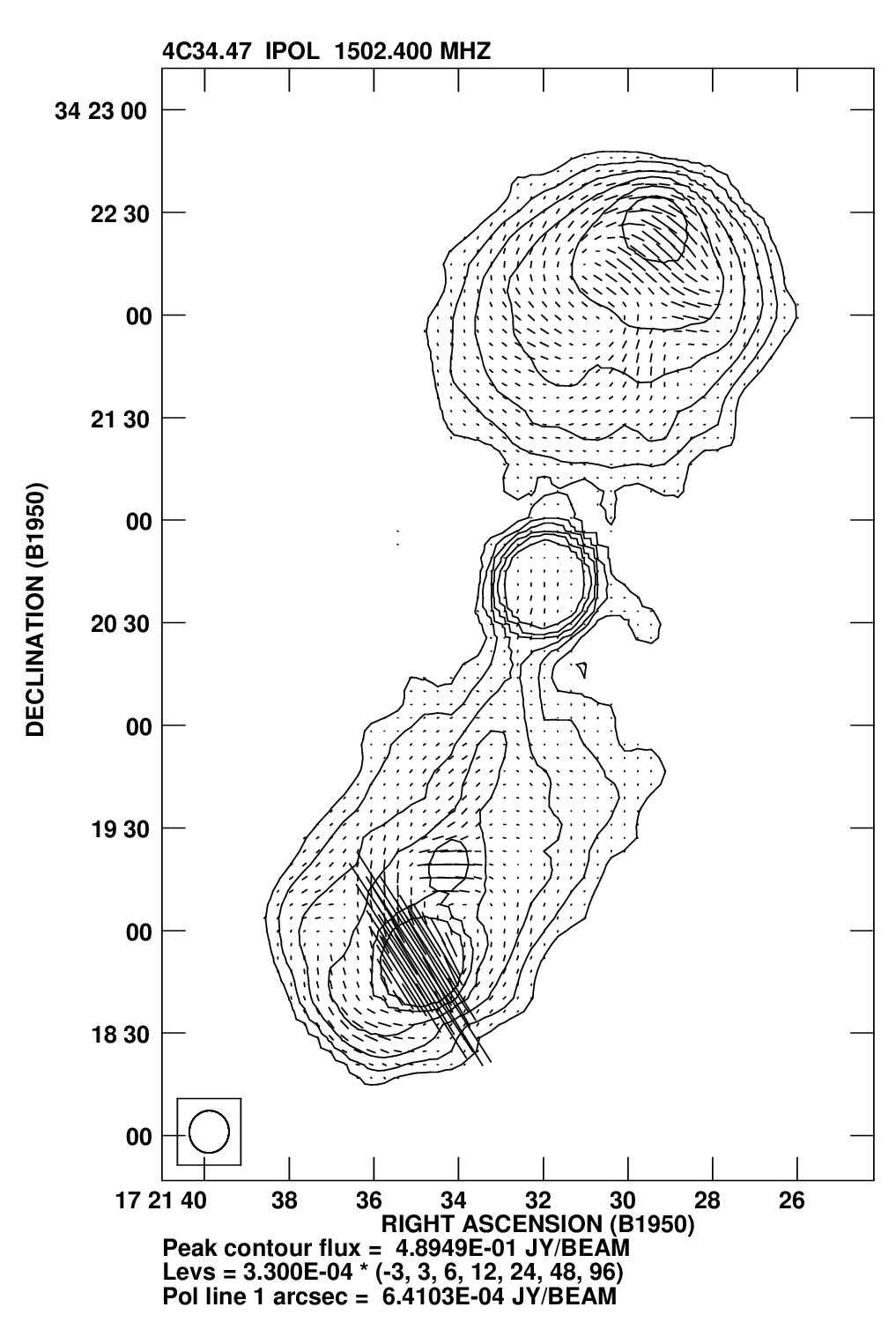}
\end{minipage} \\

\begin{minipage}{7cm}
\includegraphics[scale=0.45]{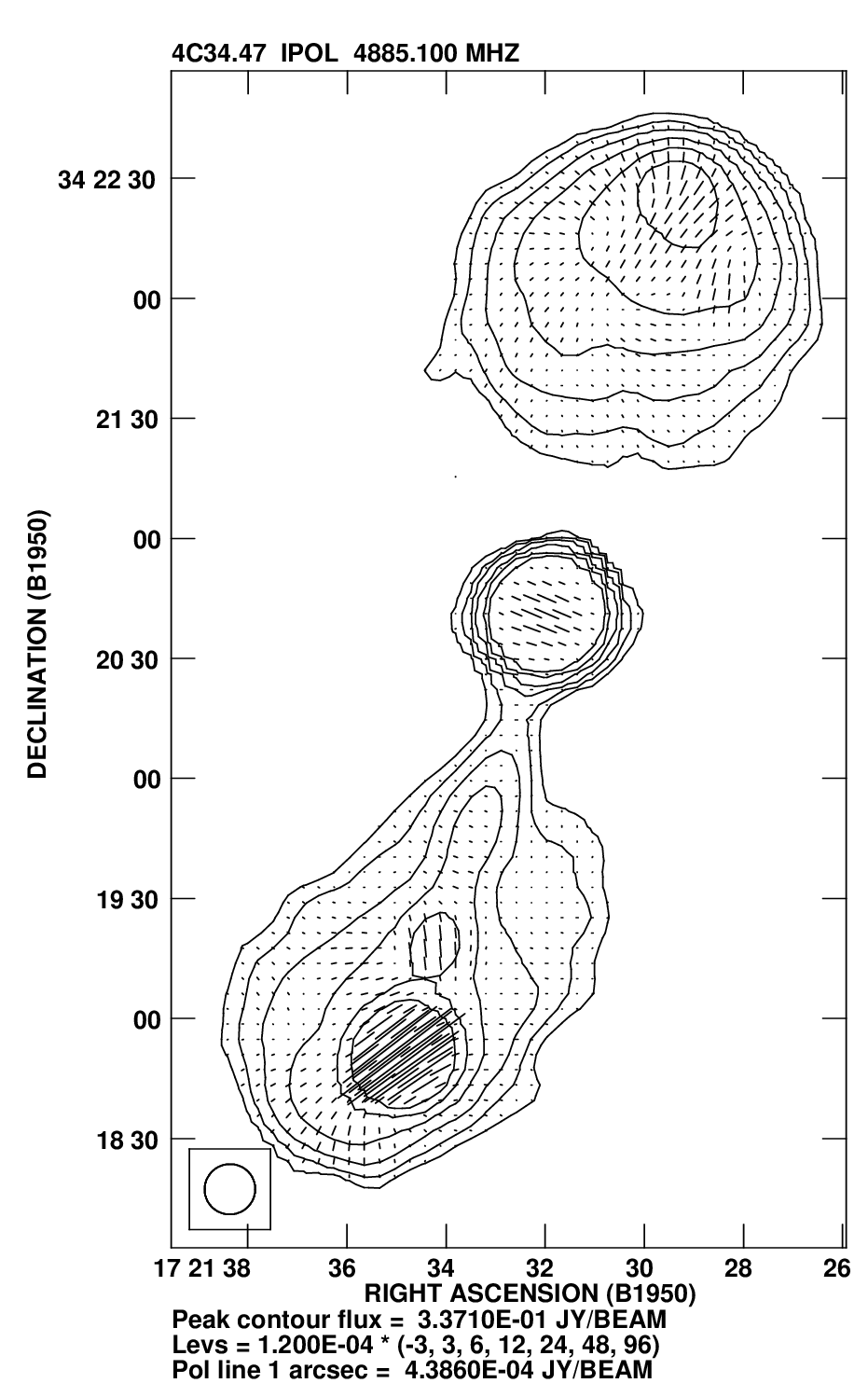}
\end{minipage} &

\begin{minipage}{7cm}
\psfrag{DECLINATION \(B1950\)}{}
\includegraphics[scale=0.45]{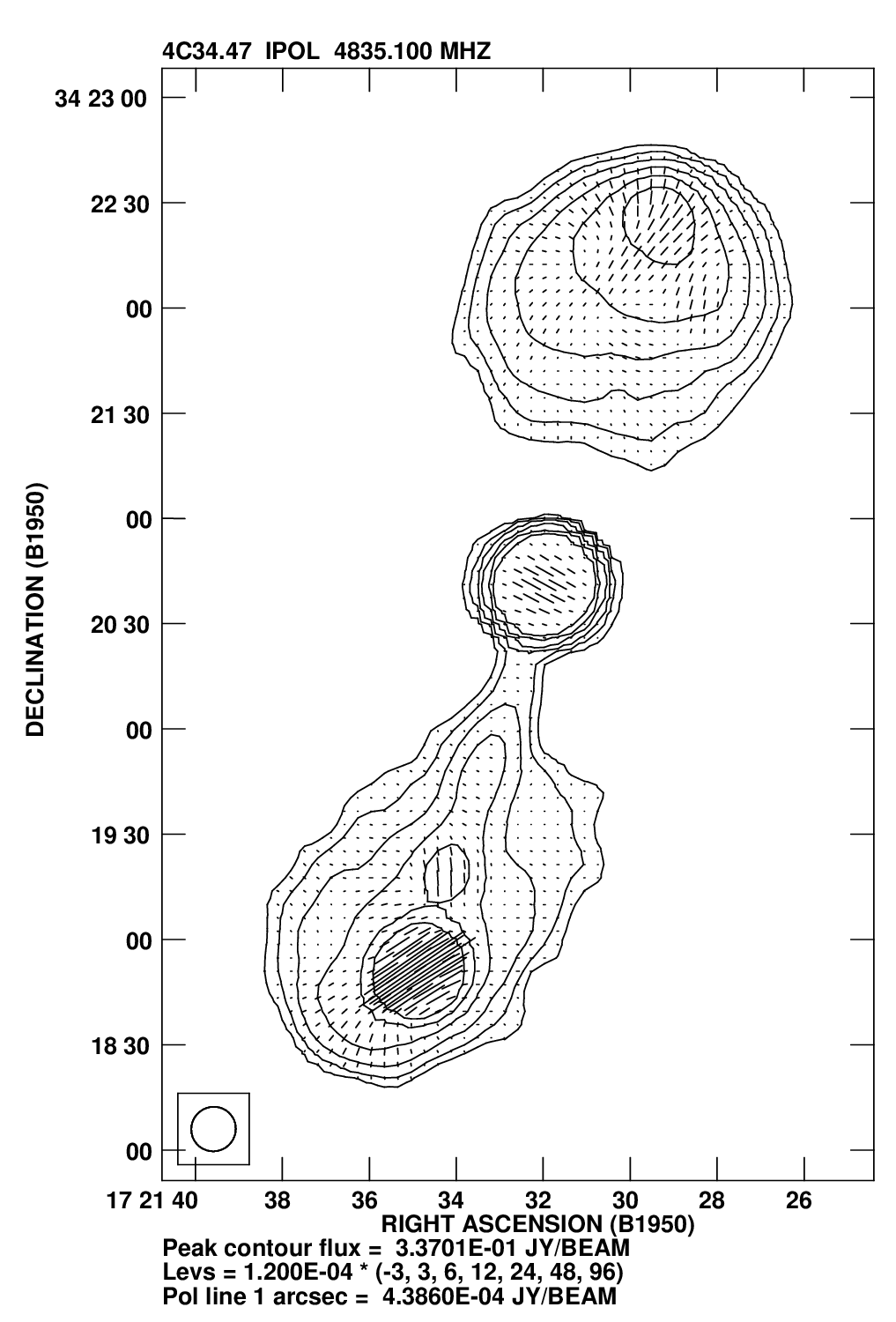}
\end{minipage} \\

\end{tabular}
\caption{Contour map of the total intensity with polarization intensity
         and E-vectors overlaid at different frequencies.
         Top left: 1452.4\,MHz; top right: 1502.4\,MHz;
         bottom left 4885.1\,MHz; bottom right 4835.1\,MHz.}
\label{fig:figure5}
\end{figure*}

\begin{figure*}[htb!]
\centering
\begin{tabular}{l l}
\begin{minipage}{7cm}
\psfrag{RIGHT ASCENSION \(B1950\)}{}
\includegraphics[scale=0.45]{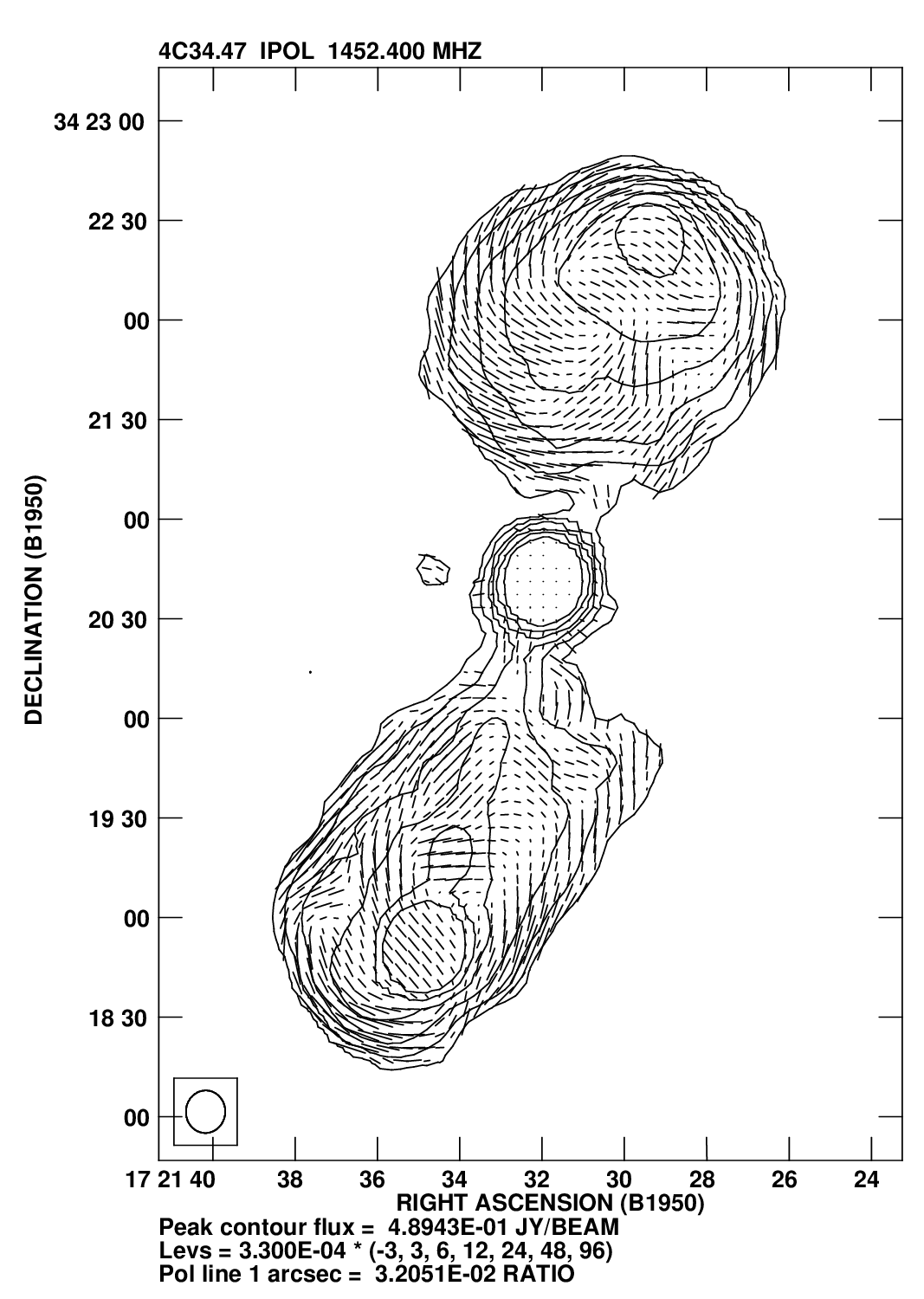}
\end{minipage} &

\begin{minipage}{7cm}
\psfrag{RIGHT ASCENSION \(B1950\)}{}
\psfrag{DECLINATION \(B1950\)}{}
\includegraphics[scale=0.45]{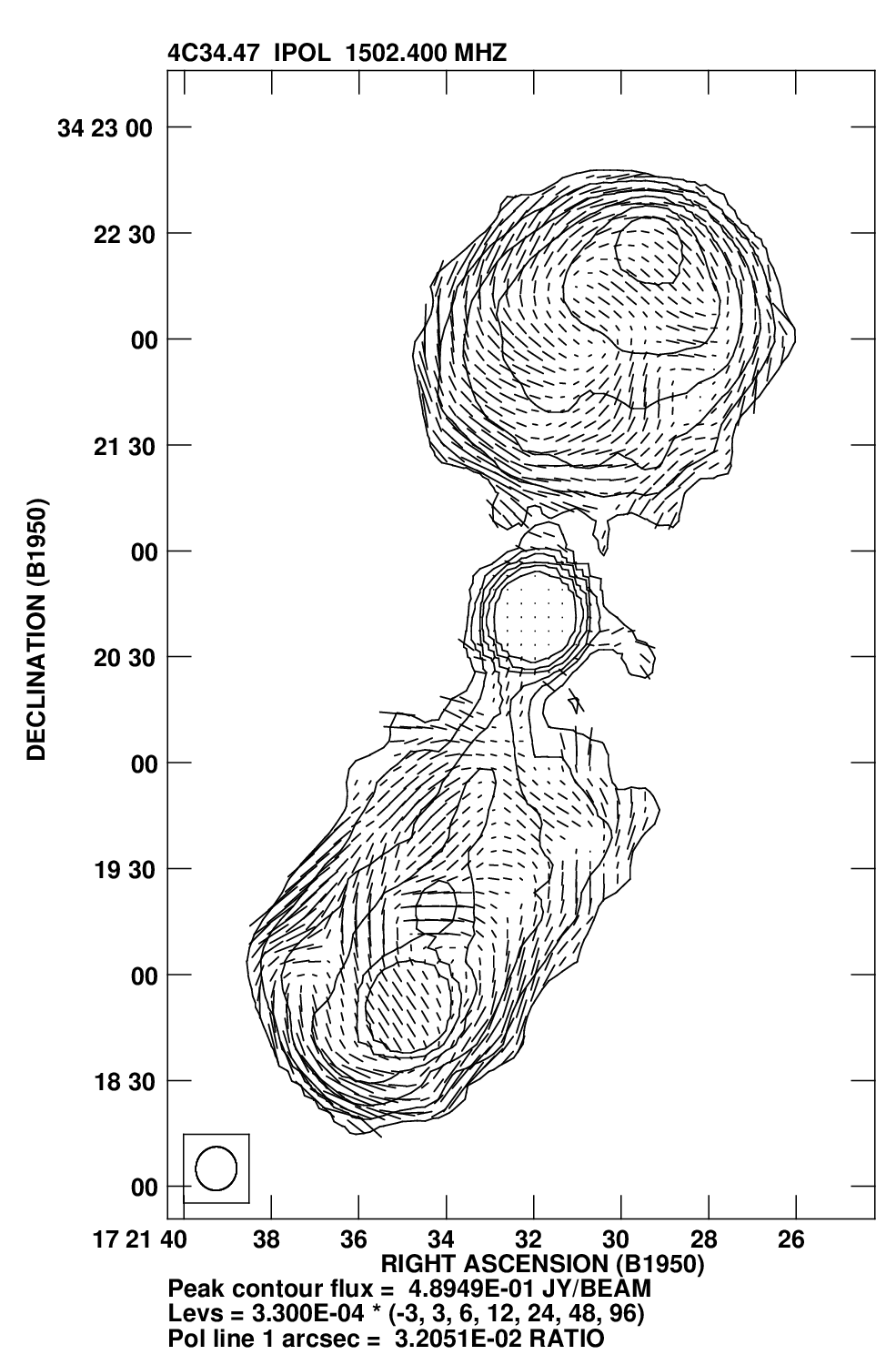}
\end{minipage} \\

\begin{minipage}{7cm}
\includegraphics[scale=0.45]{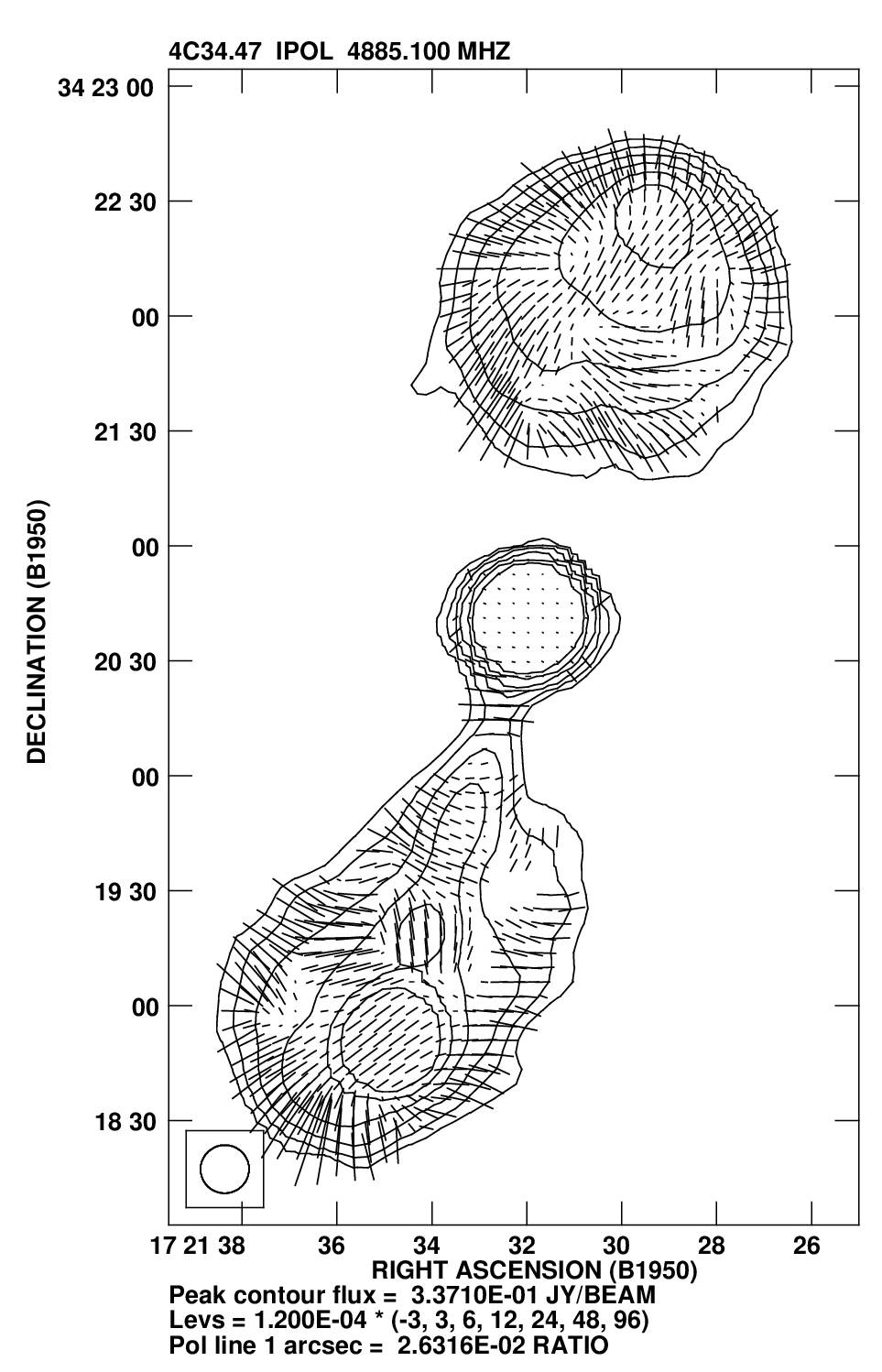}
\end{minipage} &

\begin{minipage}{7cm}
\psfrag{DECLINATION \(B1950\)}{}
\includegraphics[scale=0.45]{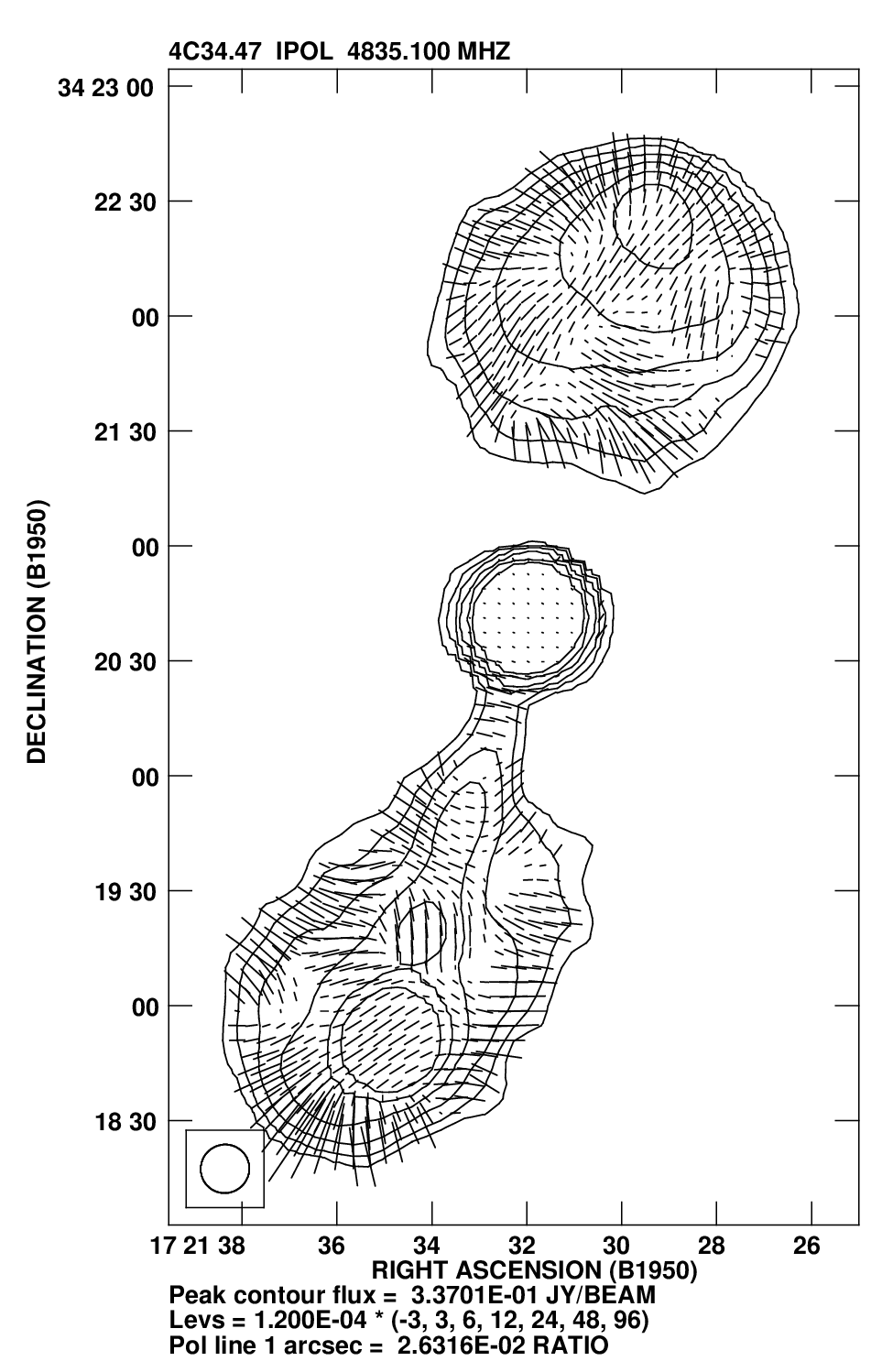}
\end{minipage} \\

\end{tabular}
\caption{Contour map of the total intensity with linear polarization
         degree and E-vectors overlaid at different frequencies.
         Top left: 1452.4\,MHz; top right: 1502.4\,MHz;
         bottom left 4885.1\,MHz; bottom right 4835.1\,MHz.}
\label{fig:figure6}
\end{figure*}

As can be readily seen, the 1.4\,GHz and 5\,GHz images display roughly
$\pi$/2 radians rotation.  However, that rotation is $n\pi$ ambiguous. 
In order to assess the magnitude and origin of the Faraday rotation, we
therefore used the two separate 1.4524 and 1.5024\,GHz IFs and the
5\,GHz data.  These frequencies permit an unambigous fit of Rotation
Measure, $RM$, values, using

\begin{equation}
\beta = RM \cdot \left( \lambda_{2}^{2} - \lambda_{1}^{2} \right),
\label{eq:eq4}
\end{equation}

\noindent
where $\beta$ is the measured angle of rotation.  The $RM$ values
inferred from these fits at various locations (see Fig.~4) are listed in
Table \ref{tab:RM}.  The southern hot spot value implies a rotation at
that position of $\sim$6$^{\circ}$ between the two 1.4\,GHz IFs -- a
value which indeed can be seen in the relevant images.  At 5\,GHz, the
rotation between the two IFs is reduced to $\sim$1$^{\circ}$.  This
implies that the measured angles of polarization at 5\,GHz trace the
source-intrinsic magnetic fields.  The source integrated Faraday
rotation measure is 41.4 rad m$^{-2}$.  This translates to a Faraday
rotation of 92{\fdg}1, from 20cm to 6cm observing wavelength.  The
rotation being roughly constant, most of it must be attributed to the
Galactic foreground.  The $RM$ value of 4C\,34.47 is moreover in line
with values for nearby objects \citep{Simard1980}.

\begin{table}[htb]
\caption{RM values at specific locations.}
\begin{center}
\begin{tabular}{|c|c|c|c|c|}
\hline
Pos. & RM (rad m$^{-2}$) & $\sigma_{RM}$ (rad m$^{-2}$) &
  Rotation $\beta$ ($^{\circ}$) & $\sigma_{\beta}$ ($^{\circ}$)\\
\hline
\hline
A & 44.51 & 6.45  & 99.05 & 14.35 \\
B & 36.88 & 4.25  & 82.07 & 9.46  \\
C & 41.60 & 3.17  & 92.57 & 7.05  \\
H & 42.47 & 10.14 & 94.51 & 22.57 \\
I & 37.75 & 4.96  & 84.01 & 11.04 \\
\hline
\end{tabular}
    \label{tab:RM}
\end{center}
\end{table}

If the Faraday rotating medium is located between the source and the
observer, then the only effect is a net rotation of the plane of
polarization.  However, if the Faraday rotating medium is mixed up with
the emitting region, then radiation emitted from different depths and
from different sight lines within the observing beam is rotated by
different amounts, thus reducing the net polarization over extended
areas. 

In order to assess the nature of the (small) Faraday depolarization,
particularly in the light of the depolarization asymmetry discussed in
Sect.~1, we have measured it at twelve different areas of the radio
source.  Eight of the selected locations are chosen to be large areas,
because larger areas will give more averaging hence more accurate
results.  We have also determined the depolarization in the hotspots, in
the core, and in the southern tip of the northern lobe (the closest
location to the core on the north side with reliable P/I detections). 
The DP value is the ratio of P/I at 5\,GHz and P/I at 1.4\,GHz.  Values
greater than unity imply depolarization; values smaller than unity
indicate re-polarization.  The measurements are tabulated in Table
\ref{tab:DP}. 

\begin{table*}[htb!]
\caption{DP values at different locations.}
\begin{center}
\begin{tabular}{|c||c|c||c|c||c|c||}
\hline
\hline
\textbf{Northern lobe} & \multicolumn{2}{|c||}{$\Pi_{6cm}$ (\%)} & \multicolumn{2}{|c||}{$\Pi_{20cm}$ (\%)} & \multicolumn{2}{|c||}{DP}       \\
                             &  IF-1  & IF-2    &   IF-1 & IF-2     & IF-1 &IF-2 \\
\hline
\hline
At the hot spot        & 9.20 & 9.80 & 9.44 & 9.58 & 0.97 & 1.02 \\
North of hot spot area & 9.63 & 9.76 & 9.56 & 9.72 & 1.01 & 1.00 \\
East-side in the lobe  & 13.05 & 12.11 & 11.94 & 11.84 & 1.09 & 1.02 \\
West-side in the lobe  & 8.98 & 8.94 & 9.17 & 9.12 & 0.98 & 0.92 \\
South-side             & 11.06 & 10.06 & 9.37 & 9.51 & 1.18 & 1.06 \\
\hline
\hline
\textbf{Southern lobe} & \multicolumn{2}{|c||}{$\Pi_{6cm}$ (\%)} & \multicolumn{2}{|c||}{$\Pi_{20cm}$ (\%)} & \multicolumn{2}{|c||}{DP}       \\
                             &  IF-1  & IF-2    &   IF-1 & IF-2     & IF-1 &IF-2 \\
\hline
\hline
At the spot spot       & 12.56 & 12.57 & 12.84 & 12.90 & 0.98 & 0.97 \\
South of hot spot area & 12.28 & 12.15 & 12.58 & 12.76 & 0.98 & 0.95 \\
East in the lobe       & 15.74 & 15.24 & 16.44 & 16.56 & 0.96 & 0.92 \\
West in the lobe       & 9.36 & 9.59 & 9.46 & 9.94 & 0.99 & 0.96 \\
Jet area               & 10.81 & 11.08 & 10.94 & 11.19 & 0.99 & 0.99 \\
\hline
\hline
\textbf{Around and at the core} & \multicolumn{2}{|c||}{$\Pi_{6cm}$ (\%)} & \multicolumn{2}{|c||}{$\Pi_{20cm}$ (\%)} & \multicolumn{2}{|c||}{DP}       \\
                             &  IF-1  & IF-2    &   IF-1 & IF-2     & IF-1 &IF-2 \\
\hline
\hline
Core & 0.91 & 1.00 & 0.20 & 0.23 & 4.48 & 4.29 \\
Southern tip of northern lobe & 14.07 & 14.77 & 10.77 & 11.46 & 1.31 & 1.29 \\
\hline
\end{tabular}
    \label{tab:DP}
\end{center}
\end{table*}

As can be readily seen, hardly any significant depolarization is
measured: except for the core and the lobe emission just north of the
core, all values are unity within the measurement errors.  This is a
remarkable result, suggesting that little ionized gas is mixed in with
the radio plasma and that little or no depolarizing medium is present
around the giant radio source. 

\section{Discussion}

\subsection{Orientation of the radio jet}

By measuring the jet-to-counterjet flux density ratio and knowing the
bulk flow velocity in the jets, the object's angle $\theta$ w.r.t.  its
line of sight can be recovered, assuming an intrinsically symmetric
object experiencing Doppler boosting resulting in its asymmetric
appearance.  Following \citet{Scheuer1979}, this phenomenon can be
quantified using Eq.  (\ref{eq:sidedness}):

\begin{equation}
\frac {S_{j}} {S_{cj}} = \left[ \frac {1 + \beta_{j} cos\theta} {1 - \beta_{j} cos\theta}    \right]^{2-\alpha}, 
\label{eq:sidedness}
\end{equation}

\noindent
where $\beta_{j}$ is the velocity of the jet in units of $c$ and
$\alpha$ the jet spectral index.  If on the other hand the angle is
known but the velocity needs to be recovered, Eq. (\ref{eq:sidedness})
can be rewritten in the form for $\beta$ cf.  Eq. (\ref{eq:sidedness2}).
Eq. (\ref{eq:sidedness3}) gives the corresponding Lorentz factor. 

\begin{equation}
\beta_{j} = \frac {1} {cos\theta} \cdot  \frac {s - 1} {s + 1}
\label{eq:sidedness2}
\end{equation}

\begin{equation}
\gamma = 1 /  \left\lbrace 1 - (1/cos^{2}\theta)[(s - 1)/(s + 1)]^{2} \right\rbrace^{\frac {1} {2}},
\label{eq:sidedness3}
\end{equation}

\noindent
where s = $(S_{j}/S_{cj})^{\frac {1} {2-\alpha}}$. \\

The average jet spectral index obtained from the results in Fig.~4 is
$-0.6$.  The high resolution 5\,GHz image indicates a lower limit for
the jet-counterjet ratio of $S_{j}/S_{cj}=24.0$, with a sigma of 2.9. 
Rewriting Eq. (\ref{eq:sidedness2}) yields an upper limit to the
angle $\theta$:

\begin{equation}
\theta = \arccos\left(\frac {1} {\beta_{j}} \cdot  \frac {s - 1} {s + 1}  \right)
\label{eq:sidedness4}
\end{equation}

\noindent

Hence, assuming an intrinsically symmetric radio sources, we infer that
the angle to the line of sight of 4C\,34.47 can not be larger than
57$^{\circ}$ (for $\beta_{j} = 1$).  This upper limit is not very
sensitive to the value of the jet spectral index $\alpha$.  The actual
bulk flow speed in the jet must be smaller than the speed of light. 
Moreover, the flux density of the approaching jet is taken as an average
over the whole jet and since the counterjet is not detected at all, the
maximum flux at any point on the approaching jet could also have been
used to get the ratio.  This would have resulted in a higher limit to
the flux density ratio.  The dependence of the flow velocity on the
angle at the adopted ratio of 24 is presented in Fig.~7.  For reasonably
relativistic speeds ($\geq0.9c$), this implies that for the adopted
ratio, the range of the angle is 4$^{\circ}$, i.e., between 53$^{\circ}$
and 57$^{\circ}$.  The lower limit of 53$^{\circ}$ is obviously not
strict: a higher value for the jet-counterjet ratio is quite plausible,
and hence a smaller aspect angle. 

\begin{figure}[htp!]
    \centering
    \includegraphics[scale=0.4]{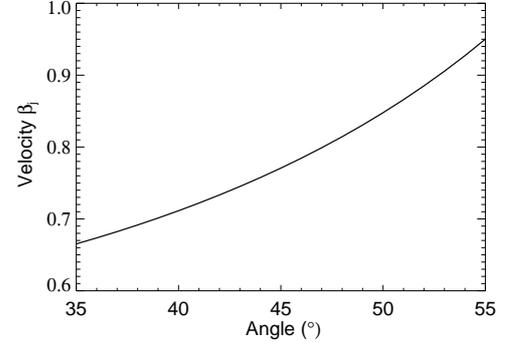}    
    \caption{Dependence of the jet bulk flow velocity on the 
     aspect angle, adopting a flux density ratio S$_{j}$/S$_{cj}$=24
     and jet spectral index $-0.6$}
    \label{fig:plot1}
\end{figure}

\subsection{Foreground Faraday rotation}

Given the roughly constant rotation of about 90$^{\circ}$ over all
components of the radio source, the Faraday rotation towards 4C\,34.47
must be mainly due to our own Galaxy.  However, the rotation is not
entirely uniform over the whole source.  As judged from Table
\ref{tab:RM}, it ranges between 80$^{\circ}$ to 100$^{\circ}$. 
\citet{Leahy1987}, analyzing foreground $RM$ behaviour on size scales of
tens of arcseconds, finds that $RM$ variations for sources within
10$^{\circ}$ of the galactic plane are $\approx$ 20~rad m$^{-2}$, while
these variations are $\approx$ 6~rad m$^{-2}$ for sources near the
galactic pole.  The galactic latitude of 4C\,34.47 is 32{\fdg}2.  Given
that they occur on size scales as small as 20\arcsec, the 4C\,34.47 $RM$
standard deviation values (Table \ref{tab:RM}) are in line with the
\citet{Leahy1987} measurements. 

\subsection{Depolarization and its asymmetry}

As presented in Sect. 3.5, there is hardly any measurable
depolarization between 6 and 20cm, except towards the core and the
southern tip of the northern lobe (about 175\,pc north of the core). 
4C\,34.47 is a very large source and these depolarization results are
therefore not entirely unexpected.  \citet{Garrington1991a} and
\citet{Garrington1991b} compare the polarization properties of 47 double
radio sources with half of them having angular dimensions in excess of
$\sim30\arcsec$, and establish a significant depolarization asymmetry for the
small sources but not for the large ones.

As \citet{Garrington1991b} show, the medium responsible for the
asymmetry is not likely to be hot ISM gas but rather the IGM because the
differences in $RM$ between lobes do not correlate with the Faraday
dispersion factor $\Delta$ --- see Eq. (\ref{eq:faradayDispersion2}). 

\begin{equation}
\Delta^{2}(x) = \int \limits_{\infty}^{x} \sigma^{2} dl
  {\rm ~~~with~~~}
  \sigma^{2} = n_{e}^{2} {B}^{2} d_{cell}
\label{eq:faradayDispersion2}
\end{equation}

From X-ray observations, the presence of a halo of hot gas has been
revealed around many galaxies and this is most likely the dominant agent
causing depolarization (which is generally observed in extragalactic
radio sources -- e.g., \citet{Strom1988}.  Within the halo gas
model, depolarization asymmetry is easily explained by the
orientation of the source, relating the visibility of a relativistic jet
to its approaching nature and the non-visibility of a (counter)jet to
its receding nature. 

Besides the orientation of the object, the size of this halo of hot gas
relative to the radio source size determines the magnitude of the
depolarization asymmetry.  The largest asymmetry should occur at the
edges of the radio source, provided the overall source is still well
contained within the hot gas halo.  4C\,34.37 obviously violates this
requirement.  In fact, \citet{Garrington1991b} conclude from their
analysis that the typical halo core diameter must be about 100 kpc. 

We therefore conclude that 4C\,34.47 has largely outgrown its depolarizing
halo but note with interest that the only occurrence of significant
off-nuclear depolarization is at the invisible counterjet-side, cf. 
standard predictions.  It would be interesting to study the
depolarization properties at decimeter wavelengths with sufficient
resolution and sensitivity.  A dedicated VLA or LOFAR project
would be required for this purpose. 

It should be noted also that the pure orientation explanation for the
depolarization behaviour is too simplistic.  The effects of host galaxy
environment as well as redshift were proven to play a role \citep{Liua,
Liub, Dennett1997, Dennett1999, Goodlet2005}.  An in-depth study by
\citet{Fernini2001} found no convincing differences of
depolarization-ratio's between classes of -- large -- radio galaxies and
quasars.  On the basis of its depolarization properties, 4C\,34.47
nevertheless fits the picture of an inclined radio source with its
northern lobe directed away from us. 

\subsection{The orientation of 4C\,34.47}

As discussed above, its strong one-sided jet together with its
polarization properties argue for an aspect angle value of the 4C\,34.47
radio source axis, $\theta << 90\degr$.  A fairly simplistic treatment
of the jet-counterjet ratio permits to determine an upper limit to the
inclination angle w.r.t.  the sight line of $\theta < 57^{\circ}$, or
$\theta << 57^{\circ}$.  That constraint would be largely consistent
with the $\theta \la 45^{\circ}$ restriction for quasars within the
quasar -- radio galaxy unification model \citep{Barthel1989b}. 

How does this orientation of the large scale radio source compare to the
orientation of the milliarcsec scale radio jet in its bright radio core?
For this jet, \citet{Barthel1989a} and \citet{Hooi1992} report a knot
proper motion of 0.29 milliarcsec/year.  This value translates to a
projected expansion speed of $3.9c$ using the here adopted cosmology,
and that apparent superluminal motion implies a maximum allowable
inclination\footnote{\citet{Barthel1989a} determined a maximum allowable
angle of 44$^{\circ}$ for 4C\,34.47, using a different cosmological
model} of 29$^{\circ}$, within the framework of the standard model for
superluminal motion \citep{Pearson1987}.  A source aspect angle $\theta
= 14^{\circ}$ would minimize the requirement on the relativistic bulk
flow speed, $\gamma$, in the small scale jet jet to $\gamma \gid 3.9$,
in order to explain the observed superluminal motion.  Such angles
however would suggest jet-counterjet flux ratio's at least one order of
magnitude larger than the lower limit measured above. 

There is no a priori reason that the small scale jet should have exactly
the same inclination as the large scale jet.  In fact,
\citet{Barthel1989a} noted a {\it projected} misalignment angle of
$5^{\circ}$ between the large and the small scale jets in 4C\,34.47. 
The inferred small scale aspect angles of 29$^{\circ}$ and 14$^{\circ}$
nevertheless would imply deprojection factors $1/{\rm sin}{\,}\theta =
2.0$ and 4.1 respectively, whereas the large scale limit $\theta \la
57^{\circ}$ would suggest a factor of at least 1.2.  Such factors
indicate a physical size of 4C\,34.47 in the range 1.0 -- 3.4\,Mpc. 
This is large, but not excessively large.  Long time record holder radio
galaxy 3C\,236, measuring 4.3\,Mpc was recently overtaken by
J1420$-$0545, having a projected size of 4.6\,Mpc \citep{Machalski2008}. 
With reference to e.g., \citet{Ishwara1999} and the Introductory
Section, double lobed radio sources associated with QSOs have projected
sizes up to $\approx$ 1.2\,Mpc, with the occasional outlier at 2\,Mpc,
whereas radio galaxies reach up to $\approx$ 2.5\,Mpc, with the
occasional outlier at 4\,Mpc.  Even seen under a small angle, 4C\,34.47
is not uncomfortably large.  That small aspect angle finds additional
support in the fact that the broad-emission line region in the
associated QSO B1721+343 is probably seen face-on \citep{Miley1979,
Wills1986}, and in the variable nature of its bright core, as discussed
above. 

On the assumption of a symmetric radio source, with the same intrinsic
core-to-hotspot distances north and south, the approaching southern
hotspot should have a larger projected distance from the core than the
receding northern one, due to the light travel time difference.  This is
indeed the case: the measured difference is 19\,kpc projected distance. 
Irregardless the exact inclination of the radio source, that difference
can only be modeled adopting a fairly low hotspot advance speed,
$\approx 0.02c$ -- $0.04c$.  Such speeds are indeed typical for the
class of low luminosity FR2 radio sources \citep{Alexander1987}, to
which 4C\,34.47 belongs, and are furthermore consistent with the absence
of a significant difference in hotspot spectral index
\citep[e.g.,][]{Ishwara2000}. 

\subsection{Giant among giants ....}

How does giant quasar 4C\,34.47 compare to other giant radio sources?
Giant radio sources are loosely defined as having a projected dimension
in excess of 1\,Mpc (by fans of an $H_0 = 50$ cosmology).  Using the
current cosmology standard, 4C\,34.47 has a projected size of 0.84\,Mpc. 
As discussed in the previous Section, it definitely classifies among the
very large radio sources associated with QSOs, and we note in passing
that the size distributions of the giant radio sources as mentioned
briefly in the previous Section are consistent with predictions from the
standard quasar -- radio galaxy unification model.  The quasars,
including the giants and obviously 4C\,34.47, stand out with higher core
fractions, which is also cf.  unification and beaming predictions
\citep[e.g.,][]{Hough1989, Ishwara2000}. 

Giant radio sources generally have relatively low radio luminosities.  A
$P$--$D$ diagram (power--size) comparing giant radio sources with the
overall extragalactic 3CR population was shown by \citet{Ishwara1999}. 
The deficit of large, high-power objects is generally explained as being
due to the effects of aging, whereby lobes in old radio sources have
lost substantial fractions of their radio power through expansion,
Inverse Compton, and radiative losses.  4C\,34.47, having a 1.4\,GHz
radio luminosity of just over $1 \times 10^{26}$ W/Hz (see Sect.  3.1),
fits that model.  Large radio lobe volumes emitting weakly suffer from
relatively large Inverse Compton losses (to the redshifted cosmic
background).  Also for 4C\,34.47, these losses dominate over the source
intrinsic radiation losses.  Inverse Compton losses limiting the age of
the radio source, giant radio sources such as 4C\,34.47 are believed to
be rare at high redshift, where the inverse Compton losses are
excessive. 

\section{Conclusions}

Sensitive, detailed, dual-frequency, multi-resolution images of the
large double-lobed radio source 4C\,34.47 associated with the $z=0.206$
QSO B1721+343 obtained with the VLA were presented.  The radio
structural as well as radio spectral and polarization properties of this
giant radio source were studied.  A prominent one-sided jet of record
length was discovered.  This jet, together with the bright, variable
radio core, provide support for the picture whereby the axis of radio
source is substantially offset from the sky plane, despite the large
projected size of the overall source.  As such, the VLA imaging supports
the preferred orientation of the radio source, as indicated earlier from
the superluminal jet observed in its radio core.  The measured large
scale radio (de)polarization properties are not in contradiction with
this picture, but suggest also that 4C\,34.47 has largely outgrown its
magnetoionic halo, again cf.  expectation.  4C\,34.47 is a slowly
expanding, old FR2 radio source, with its jet axis seen under a small
aspect angle causing relativistic beaming phenomena to be observed. 
Otherwise, 4C\,34.47 is not different from other giant radio sources
oriented differently. 

\begin{acknowledgements}

The authors acknowledge careful reading by the referee.  SH acknowledges
expert advice from Robert Laing and Richard Strom, during their visits
to the Kapteyn Astronomical Institute.  PDB acknowledges the NRAO VLA
staff for expert observing and assistance in data handling and Wil van
Breugel for his early interest in this project.  The National Radio
Astronomy Observatory is a facility of the National Science Foundation
operated under cooperative agreement by Associated Universities, Inc. 

\end{acknowledgements}

\bibliographystyle{aa}
\bibliography{giant.bib}

\end{document}